\documentclass[a4paper,pra,twocolumn]{revtex4}  %% REVTeX 4.0
\usepackage[latin1]{inputenc}
\usepackage[T1]{fontenc}
\usepackage[english]{babel}
\usepackage{moreverb}
\usepackage{fancyhdr}
\usepackage{graphics}
\usepackage{SIunits}
\usepackage{amsmath}
\usepackage{epsfig}
\usepackage[small]{subfigure}
\newcommand{\jh}[0]{\hat{J}}

\newcommand{\sh}[0]{\hat{S}}
\newcommand{\ks}[0]{\kappa^2}
\newcommand{\kst}[0]{\tilde{\kappa}^2}

\newcommand{\gsd}[0]{\gamma^{sd}}

\begin{document}
\title{Entanglement of large atomic samples: a Gaussian state
  analysis} \author{Jacob Sherson and  Klaus M\o lmer} \affiliation{QUANTOP,
  Danish Research Foundation Center for Quantum Optics, Department of
  Physics and Astronomy, University of Aarhus, DK 8000 Aarhus C,
  Denmark} \email{sherson@phys.au.dk,moelmer@phys.au.dk}

\begin{abstract}
  We present a Gaussian state analysis of the entanglement generation
  between two macroscopic atomic ensembles due the continuous probing
  of collective spin variables by optical Faraday rotation.  The
  evolution of the mean values and the variances of the atomic
  variables is determined and the entanglement is characterized by the
  Gaussian entanglement of formation (GEoF) and the logarithmic
  negativity.  The effects of induced opposite Larmor rotation of the
  samples and of light absorption and atomic decay are analyzed in
  detail.

\end{abstract}

%\ocis{}

\maketitle   %%  REQUIRED FOR REVTeX 4.0

\section{Introduction}
\label{sec:intro}

Macroscopic samples of atoms as a resource of entanglement have
attracted attention because of their robustness to single particle
losses which leads to msec lifetimes of the entangled states and
because of the effective coupling to light as needed for quantum
repeaters and memories in quantum communication networks
\cite{Kimble,mynature04}. The theoretical proposal \cite{Duan,kuzmichspinsq}
surprisingly showed that by merely probing the state of atomic samples
with light from a classical light source, one induces an atomic
dynamics where the quantum state evolves by state reduction to
entangled states. The experimental implementation of the proposal
\cite{Duan} led to the first demonstration of entanglement between
macroscopic ($\sim 10^{12}$) numbers of atoms \cite{nature,distent}.
In this work we extend the theoretical Heisenberg picture analysis in
\cite{Duan} with an analysis addressing directly the quantum state of
the atoms and its time evolution due to the interaction with the
continuous wave (cw) probe field, the back action of the measurements,
obtained continuously in time, and light absorption and atomic decay.
We note that a quantum trajectory approach with simulated state vector
dynamics was presented in \cite{Dilisi,Dilisi2} to provide a
microscopic description of the dynamics, but because of the dimensions
of the Hilbert spaces involved, these simulations were restricted to a
few tens of atoms.  We retain in this work the careful attention of
\cite{Dilisi,Dilisi2} to the quantum mechanical effects of the
measurement in a treatment of macroscopic samples by a practically
exact Gaussian Ansatz for the quantum states. This permits the use of
the powerful formalism of correlation matrices for Gaussian states
\cite{Eisert,GiedkeCirac}.

The successful experimental verification of entanglement of
macroscopic samples \cite{nature} utilizes Larmor rotation of the
samples. This presents an experimental advantage compared to
measuring on only one EPR quadrature at a time, but until now
there has been no thorough theoretical examination of the precise
effect of these rotations on the entanglement generation rate. We
give explicit analytic expressions for the entanglement generation
rate as a function of rotation frequency in the absence of light
absorption and numerical results in the presence of light
absorption and atomic decay.

In section \ref{sec:int} we introduce the description of the atomic
and light variables and briefly discuss the basic interaction. In
section \ref{sec:nonoise} we introduce a Gaussian description of the
interaction and solve a nonlinear differential equation for the atomic
variables in the absence of light absorption. The entanglement is
quantified in terms of the GEoF. In section \ref{sec:noise} we
describe the effects of light absorption and we present analytic
solutions for the atomic variables in the case of small decoherence
effects and numerical solutions for the general case. In section
\ref{sec:meanvalues} we discuss the evolution of the mean values of
the atomic variables during the interaction. Section
\ref{sec:conclusion} concludes the paper.
 
\section{Setup and Interaction}
\label{sec:int}

\begin{figure}[h]
\includegraphics[width=0.40\textwidth]{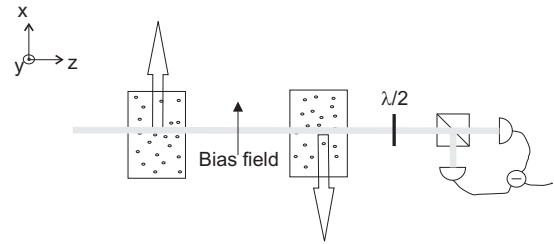}
\caption{\small 
  A continuous wave light beam linearly polarized along the x-axis is
  sent through two macroscopic samples of atoms optically pumped with
  collective spins in the positive and the negative x-direction
  respectively. The polarization rotation of the field is monitored
  continuously. A bias field along the x-direction is applied to
  induce Larmor rotation of the atomic spins in the y-z plane during
  the measurement. }
 \label{fig:twocells}
\end{figure}

We consider the system studied experimentally in \cite{nature,distent}
and sketched in Fig. \ref{fig:twocells} with two macroscopic samples
of spin 1/2 particles polarized along the positive and negative x-axis
respectively.  The samples interact with an off resonant linearly
polarized light beam giving rise to a Kerr-interaction between the
macroscopic spin operator, $\vec{J}$, and the Stokes operator of
light, $\vec{S}$. We assume the two atomic samples to be prepared
close to the maximally polarized state along x with magnitudes
$J_{x1}=J_x\equiv N_{a}/2$ and $J_{x2}=-J_x$, where $N_a$ is the
common number of atoms in each of the two samples. Here and in the
remaining part of this paper we set $\hbar=1$. For macroscopic samples
of atoms the quantum mechanical uncertainty in $J_x$ is negligible
compared to the magnitude of $J_x$ which can therefore be considered a
classical number. With large number of photons in the probing beam the
same argument applies to the Stokes vector component $S_x$.  Defining
a vector of observables
\begin{equation}
\mathbf{y}=\left(\begin{array}{c}
x_{A1}\\
p_{A1}\\
x_{A2}\\
p_{A2}\\
x_{L}\\
p_{L} \end{array}
 \right)
 =
\left(\begin{array}{c}
J_{y1}/\sqrt{J_x}\\
J_{z1}/\sqrt{J_x}\\
-J_{y2}/\sqrt{J_x}\\
J_{z2}/\sqrt{J_x}\\
S_{y}/\sqrt{S_x}\\
S_{z}/\sqrt{S_x} \end{array}
 \right)
\end{equation}
the system is described to a good approximation by operators which
obey the usual position and momentum commutation relation
$[x_i,x_j]=[p_i,p_j]=0,[x_i,p_j]=i \delta_{ij}$.  The Hamiltonian for the
interaction between the light and either of the two samples will be
given by an expression of the form $H_i=\kappa p_i p_L$. The coupling
strength $\kappa$ is proportional to the square root of the number of
atoms and the square root of the photon number, and it will be related
to other physical parameters in the numerical examples presented
below.

To model the continuous interaction between the atoms and the incoming
cw-light field we propose along the lines of \cite{magnetolbm} to
split the light field into independent slices of duration $\tau$. The
interaction between the samples and each light segment are then
treated one after the other. The continuous interaction and detection
of the resulting field then corresponds to taking the $\tau \to 0$
limit.

Since both the number of photons and atoms are very large and the
initial polarized state of the atoms and the light is a minimum
uncertainty state, a Gaussian distribution function for the quantum
variables is valid. This form is preserved both by the interaction
and by the detection \cite{Eisert,GiedkeCirac} so we can use the powerful
formalism of correlation matrices to describe the dynamical
evolution of the system. Within the Gaussian approximation all
information is contained in the first two moments of the quantum
variables. We are interested in the entanglement properties of the
samples which are not changed by local displacement operations so
the second moments are of primary interest. These are collected in
the 6x6 covariance matrix defined by $\gamma_{ij}=2
Re\left<(y_i-\left<y_i\right>)(y_j-\left<y_j\right>)\right>$.
Knowing how the covariance matrix is updated during the
interaction and by the detection allows us to monitor the dynamics
real-time.

\section{No light induced decoherence}
\label{sec:nonoise}

We model the evolution of the atomic system from $t$ to $t+\tau$ by
taking sequentially into account the interaction of initially coherent
light with each atomic sample, the rotation of the samples, and the
homodyne detection of the $x_L$ quadrature of the light. The evolution
of $\mathbf{y}$ in the Heisenberg picture due to the interaction
between the light segment and samples 1(2) is given by
$\mathbf{y}(t+\tau)=S_{1(2)}\mathbf{y}(t)$ with the interaction
matrices:
\[ S_1=\left( \begin{array}{cccccc}
1 & 0 & 0 & 0 & 0 & \kappa_\tau  \\
 0 & 1 & 0 & 0 & 0 & 0 \\
 0 & 0 & 1 & 0 & 0 & 0 \\
 0 & 0 & 0 & 1 & 0 & 0 \\
 0 & \kappa_\tau    & 0 & 0 & 1 & 0 \\
 0 & 0 & 0 & 0 & 0 & 1
\end{array} \right)
\qquad S_2=\left( \begin{array}{cccccc}
1 & 0 & 0 & 0 & 0 & 0 \\
 0 & 1 & 0 & 0 & 0 & 0 \\
 0 & 0 & 1 & 0 & 0 & \kappa_\tau \\
 0 & 0 & 0 & 1 & 0 & 0 \\
 0 & 0 &0&\kappa_\tau    & 1 & 0 \\
 0 & 0 & 0 & 0 & 0 & 1
\end{array} \right)
 \]
Taking into account also the Larmor precession and the detection,
 the correlation matrix will evolve according to:
\begin{equation}
  \label{eq:nonoise}
  \gamma(t+\tau)=M[R\cdot S_2 \cdot S_1 \cdot \gamma(t) \cdot
S_1^T\cdot S_2^T \cdot R^T]
\end{equation}
where R denotes a block diagonal matrix rotating the atomic variables
of the samples an angle $\pm\omega \tau$  and leaving the
light variables unchanged. M[...] denotes the effect of the homodyne
detection. Let the covariance matrix before the homodyne measurement
be given by:
\begin{equation} \gamma=\left( \begin{array}{cc}
\gamma_a&\gamma_c\\
\gamma^T_c &\gamma_b
\end{array} \right)
\label{eq:gamalgeneral}
\end{equation}
where $\gamma_a$ is a 4x4 matrix describing the atomic subsystem,
and $\gamma_b$ is a 2x2 matrix describing the light system. All
atom-light correlations are contained in $\gamma_c$. After the
detection the atomic part of the correlation matrix is then given
by \cite{Eisert,GiedkeCirac}:
\begin{equation}
  \label{eq:homodet}
\gamma_a\to \gamma_a-\gamma_c(\pi \gamma_b \pi )^{-}\gamma^T_c
\end{equation}
where $\pi=\mathrm{diag}(1,0)$ because one quadrature of the light
is assumed to be detected perfectly and $(\cdot)^-$ denotes the
Moore-Penrose pseudo-inverse of a matrix.

When Eq.(\ref{eq:nonoise}) is evaluated for short time segments
$\tau$, the change in $\gamma$ is quadratic in $\kappa_\tau$.
$\kappa_\tau^2$ is proportional to the photon number in the beam
segment, i.e. proportional to $\tau$ and rewriting
$\kappa^2_\tau=\tilde{\kappa}^2 \tau$, the differential limit for
the atomic correlation matrix can be formed:
\begin{equation}
  \label{eq:dif}
\frac{d\gamma_{a}}{dt} = \mathbf{r} \gamma_{a} + \gamma_{a}\mathbf{r}^T +\tilde{\kappa}^2 (\tilde{A}-\gamma_{a} \tilde{B} \gamma_{a}^T)
\end{equation}
where:
\begin{equation} \tilde{A}=\left( \begin{array}{cccc}
1 & 0 & 1 & 0  \\
 0 & 0 & 0 & 0  \\
 1 & 0 & 1 & 0  \\
 0 & 0 & 0 & 0
\end{array} \right)
,~~ \tilde{B}=\left( \begin{array}{cccc}
0 & 0 & 0 & 0  \\
 0 & 1 & 0 & 1  \\
 0 & 0 & 0 & 0  \\
 0 & 1 & 0 & 1
\end{array} \right)
 \end{equation}
and
\begin{equation} \mathbf{r}=\left( \begin{array}{cccc}
0 & \omega & 0 & 0  \\
 -\omega & 0 & 0 & 0  \\
 0 & 0 & 0 & -\omega  \\
 0 & 0 & \omega & 0
\end{array} \right)
\label{eq:r}
 \end{equation}

 We note that the evolution of the atomic
 covariance matrix caused by the measurements given by Eq.
 (\ref{eq:homodet}) is deterministic (despite the random outcomes of the detection)
 and non-linear.

\subsection{Ricatti equation and solution}

The nonlinear differential Eq. (\ref{eq:dif}) can be solved using the Ricatti
method as eg. mentioned in the appendix of \cite{Stockton}. The
generic Ricatti equation is:

\begin{equation}
  \label{eq:genric}
\frac{d\mathbf{V}}{dt} =  \mathbf{C}- \mathbf{D} \mathbf{V}(t)- \mathbf{V}(t) \mathbf{A}- \mathbf{V}(t) \mathbf{B} \mathbf{V}(t)
  \end{equation}

Using the decomposition
$\mathbf{V}(t)=\mathbf{W}(t)\mathbf{U}^{-1}(t)$ it can be shown that
the nonlinear differential equation can be replaced by the linear equation:

\begin{equation}
  \label{eq:genlinric}
\left( \begin{array}{c}
\frac{d\mathbf{W}(t)}{dt}  \\
\frac{d\mathbf{U}(t)}{dt}
\end{array} \right)
=\left( \begin{array}{cc}
-\mathbf{D} &\mathbf{C}  \\
\mathbf{B} & \mathbf{A}
\end{array} \right)
\left( \begin{array}{c}
\mathbf{W}(t)  \\
\mathbf{U}(t)
\end{array} \right)
\end{equation}

Matching our equation to the generic Ricatti equation and observing
that $\gamma_a^T=\gamma_a$ we obtain the linear set of equations

\begin{equation}
  \label{eq:ourric}
\left( \begin{array}{c}
\frac{d\mathbf{W}(t)}{dt}  \\
\frac{d\mathbf{U}(t)}{dt}
\end{array} \right)
=\left( \begin{array}{cc}
\mathbf{r} &\tilde{\kappa}^2\mathbf{\tilde{A}}  \\
\tilde{\kappa}^2\mathbf{\tilde{B}} & \mathbf{r}
\end{array} \right)
\left( \begin{array}{c}
\mathbf{W}(t)  \\
\mathbf{U}(t)
\end{array} \right)
\end{equation}
where we have used that  $\mathbf{r}=-\mathbf{r}^T$.

Choosing the $\mathbf{W}$ and $\mathbf{U}$ matrices to start out as
4x4 identity matrices this system of coupled linear differential
equations can be solved. The result is fairly complicated but can be
simplified by applying a time dependent rotation of $\mp \omega t /2$
to sample one and two respectively. A further simplification can be
made by noting that the measured quadratures are really the sum of p's
and the difference of x'es. In the sum/difference basis:
\begin{equation}
  \label{eq:sumdiffbasis}
   \left( \begin{array}{cccc}
1 &1 & 0 & 0  \\
 0 & 0 & 1 & 1  \\
 1 & -1 & 0 & 0  \\
 0 & 0 & 1 & -1
\end{array} \right)
\end{equation}
where the first and the third columns are the basis vectors
corresponding  to the sum of x'es and p's and the second and the
fourth correspond to the difference, we get the sum/difference
correlation matrix:
\begin{equation}
\gamma_a^{sd}=
\left(
\begin{array}{cccc}
 a_+ & 0 & 0
  & 0 \\
0 & \frac{1} {a_-} & 0 & 0 \\
 0 & 0 & \frac{1} {a_+} & 0 \\
 0 & 0 & 0 & a_-
\end{array}
\right)
\label{eq:gamstdformsumdif}
\end{equation}
where $a_\pm= 1 + \tilde{\kappa}^2
t\pm\frac{\tilde{\kappa}^2}{\omega} \sin (\omega t)$.  Defining
the total accumulated interaction at time $t$ as
$\kappa^2_t\equiv\tilde{\kappa}^2t$ and the total rotated angle
$\theta\equiv\omega t$ we get the main result of this section:
\begin{equation}
  \label{eq:a+-}
a_\pm= 1 + \kappa^2_t \pm\frac{\kappa^2_t}{\theta} \sin (\theta)
\to \left\{ \begin{array}{cc}
1+\kappa^2_t \pm \kappa^2_t & \theta \to 0\\
1+ \kappa^2_t &\theta \to  \infty
\end{array}\right.
\end{equation}

We see that as $\theta\to0$ we get $1+2\kappa^2_t$ and 1, i.e. the
characteristic squeezing and anti-squeezing of the measured and
their conjugate variables and no change in the unobserved ones.
For $\theta\to\infty$, i.e. after many rotations we squeeze the
two quadratures symmetrically and anti-squeeze their conjugate
variables by the factor $1+\kappa^2_t$. Note that the reduced
squeezing by a  factor of two in the rotated case comes from the
fact that we effectively only spend half the time measuring on
each quadrature. The result without rotations matches that of
\cite{Duan} and the strongly rotated result agrees with
calculations from \cite{distent}.

\subsection{Gaussian Entanglement of Formation}

As an entanglement measure we choose the recently proposed
Gaussian Entanglement of Formation(GEoF) of \cite{Giedke}. This
measure agrees with the Von Neuman entropy for pure states and it
can easily be calculated from the covariance matrix. It is given
by:
\begin{equation}
GEoF(\Delta)=c_+(\Delta)\log_2[c_+(\Delta)]-c_-(\Delta)\log_2[c_-(\Delta)]
\label{eq:geof}
\end{equation}
where $c_\pm(\Delta)=(\Delta^{-1/2}\pm\Delta^{1/2})²/4$ and $  \Delta^2=\mathrm{Var}(x_1-x_2)\mathrm{Var}(p_1+p_2)$.
Note that the small $\Delta$ approximation:
\begin{equation}
  \label{eq:geoflimit}
GEoF(\Delta)\approx \log_2\left(\frac{1}{\Delta}\right)+\frac{1}{\ln2}-2
\end{equation}
shows an error of $10^{-5}$ at $\Delta=1/100$, 0.001 at
$\Delta=1/10$, and only 1\% at $\Delta=1/5$
 so it is widely applicable.  From
Eqs. (\ref{eq:gamstdformsumdif}) and (\ref{eq:a+-}) $\Delta^2$ can be
shown to be
\begin{equation}
  \label{eq:delta}
  \Delta^2=\frac{1}{(1+\kappa^2_t)^2-\frac{\kappa^4_t}{\theta^2} \sin^2 (\theta)}
\end{equation}
which tells us that the entanglement will scale as
$\log_2(\kappa^2_t )=2\log_2(\kappa_t)$ with rotation and with
$\log_2(\kappa_t)$ without rotations. The factor of two is exactly
what one should expect since rotations enable us to squeeze both
quadratures compared to squeezing only one of them. This effect
was also observed in \cite{Dilisi}.

\begin{figure}[h]
\includegraphics{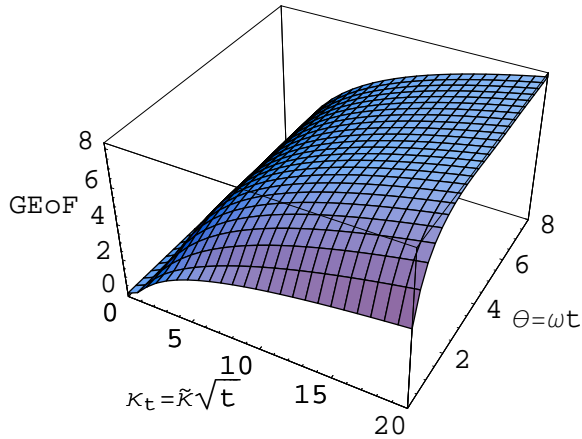}
\caption{\small
  3D plot of GEoF as a function of the accumulated interaction
  strength $\kappa_t$ and the rotated angle $\theta$.  The plot
  clearly shows that the transition from the static to the rotated
  regime has occurred already before one full revolution of the atomic
  spins has taken place.  In both the static and the rotated regions
  we clearly see the logarithmic behavior of the GEoF as a function of
  $\kappa_t$.  }  \label{fig:geof3D}
\end{figure}

In Fig. \ref{fig:geof3D} we show the entanglement plotted as a
function of $\kappa_t$ and $\theta$. As can be seen, the
transition from the static to the rotated regime occurs before one
full revolution is reached. That is, given a total interaction
time, $t$, there is no real gain in choosing the frequency larger
than $\omega_{\mathrm{crit}}=1/t$. This arises from the fact that
we measure $x_-\cos(\theta)+p_+\sin(\theta)$ and this operator
commutes with the operators measured at all previous times (other
values of $\theta$). It is therefore the accumulated measurement
on each quadrature that counts and this will not benefit from many
rotations compared to a single rotation. In the experiments of
\cite{nature,distent} $\nu_L\approx320\kilo\hertz$ and
$T_{\mathrm{probe}}\approx1\milli \second$ so the spins will
rotate several hundreds of times in one pulse and the results are
thus firmly obtained in the $\omega$ independent regime.  Note
that both axes in Fig. \ref{fig:geof3D} represent time dependent
quantities.

\begin{comment}
The form of Eq. (\ref{eq:delta}) is not ideal for studying the time
resolved entanglement generation since both variables are time
dependent.
Defining the time independent variable
$\beta\equiv\frac{\tilde{\kappa}^2}{\omega}$ we get
\begin{equation}
  \label{eq:deltanew}
  \Delta^2=\frac{1}{(1+\kappa^2)^2-\beta^2 \sin^2 (\frac{\kappa^2}{\beta})}
\end{equation}
$\kappa^2$ can now be thought of as a scaled time variable and the time
dynamics can be studied for different choices of $\beta$. This can be
seen in Fig. \ref{fig:entvskappa}. For $\beta\gg\kappa^2$
corresponding to $\omega t\ll 1$ we effectively measure a single
quadrature, $\Delta^2=1/(1+2\kappa^2)$, whereas for long timescales
the term involving $\beta$ can be ignored and we get the optimum
$\Delta^2=1/(1+\kappa^2)^2$. We now see from Fig.
\ref{fig:entvskappa} that all the different $\beta$-curves will
eventually reach the optimum curve and this will happen in the
vicinity of $\kappa=\sqrt{\beta}$ corresponding to $\omega t=1$. This
is of course very consistent with the prediction of Fig. \ref{fig:geof3D}

\begin{figure}[ht]\centering
  \includegraphics{geof_vs_kappa.eps}
  \caption{\small
    GEoF vs $\kappa^2$ for $\beta=10$(full), $\beta=50$(dashed), and
    $\beta=100$(dot-dashed). We clearly see that the curve has a kink
    around $\kappa=\sqrt{\beta}$ and after that the curve is
    independent of $\beta$ as should be expected from Eq.
    (\ref{eq:deltanew}) }
\label{fig:entvskappa}
\end{figure}

\end{comment}

\section{Light absorption and atomic decay}
\label{sec:noise}

Due to the interaction with the light field every atom has a rate by
which it is excited and decays by spontaneous emission to any of the
atomic ground states, and every photon is absorbed with a given
probability. This atomic depumping parameter and the photon absorption
probability are given by $\eta_\tau\equiv \eta \tau =\Phi \tau
\frac{\sigma}{A}\left(\frac{\Gamma}{\Delta}\right)^2$, and
$\epsilon=N_a\frac{\sigma}{A}\left(\frac{\Gamma}{\Delta}\right)^2$
respectively.  $\Phi$ is the photonic flux, $A$ is the cross section
of the atomic sample illuminated by the light, $\Delta$ is the
detuning from resonance, $\sigma$ is the cross section on resonance
for the probed transition, and $\Gamma$ is the corresponding
spontaneous decay rate. The optical density on resonance is
$\alpha_0=N_{a}\frac{\sigma}{A}$. This gives the relation for the
coupling parameter introduced above,
$\kappa^2_\tau(0)=\eta_\tau\alpha_0$. The decay of the mean spin makes
the coupling constant time dependent,
$\kappa^2_\tau=\kappa^2_\tau(0)e^{-\eta t}$. We will in our numerical
simulations use the experimentally motivated values of 5 MHz for the
decay rate $\Gamma$ and 1000 MHz for the detuning $\Delta$.  Many
results will be presented as a function of $\alpha_0$ so a brief
discussion of the experimentally realizable optical densities is in
order. In the experiments of \cite{nature,distent} $\alpha_0\approx5$
whereas optical densities of 100 or more can be routinely achieved in
MOTs (see e.g.  \cite{Ketterle}).  Using a Bose-Einstein condensate
$\alpha_0\approx1000$ can be achieved.

\subsection{Covariance matrix update and Ricatti equation}

We now derive an expression for the time evolution of the
covariance matrix due to photon loss and atomic decay. Due to atomic decay,
during a time interval $\tau$, a fraction $\eta_{\tau}$ of the
atoms decays into a random mixture of the ground states, giving
rise to the new value of the variance of one of the atomic
collective spin component:
\begin{eqnarray}
 <J_z'^2>&=&(1-\eta_{\tau})^2<J_z^2>+
(1-(1-\eta_{\tau})^2)N/4\nonumber \\
&\approx&(1-\eta_\tau)^2<J_z^2>+
\eta_\tau N/4 +\eta_\tau N/4 \label{eq:szdecay}
\end{eqnarray}
in the $\eta_\tau\ll1$ limit.  The atomic decay leads to a corresponding
reduction of the mean spin $J_x'=(1-\eta_{\tau})J_x$.

At this stage the gas contains two components: the atoms which have
not decayed, described by the first two terms in the latter expression and
the ones which have decayed, described by the last term. If nothing
else happens to the atoms, subsequent interaction with the light can
have no further effect on the random component, but a new fraction of
atoms will be randomized and we obtain the iterated expression for the
spin variance

\begin{eqnarray}
<J_{z,1}^2>&=&(1-\eta_\tau)^2<J_{z,0}^2>+ \eta_\tau \frac{N}{4} + \eta_\tau \frac{N}{4}\nonumber\\
<J_{z,2}^2>&=&(1-\eta_\tau)^2[(1-\eta_\tau)^2<J_{z,0}^2>+ \eta_\tau \frac{N}{4}]\nonumber\\
&&+ \eta_\tau \frac{N(1-\eta_\tau)}{4}+ \eta_\tau[1+(1-\eta_\tau)] \frac{N}{4}\nonumber\\
...&=&...\nonumber\\
 <J_{z,n}^2>&=&(1-\eta)^{2n}<J_{z,0}^2>\nonumber\\
&&+\eta_\tau\left[(1-\eta_\tau)^{n-1}+\sum_{j=n}^{2(n-1)}(1-\eta_\tau)^j\right] \frac{N}{4} \nonumber\\
&&+
 \eta_\tau\sum_{j=0}^{n-1}(1-\eta_\tau)^j \frac{N}{4}.\label{eq:szi}
\end{eqnarray}
In the limit $n\to\infty$ only the last term, representing the fully
random component of the gas, will contribute. The geometrical
series gives $1/\eta_\tau$, so we see that the initial squeezing will
decay exponentially as expected and we will end up with N unpolarized
atoms each contributing 1/4 to the variance as expected. Note that in
the spin 1/2 case this coincides with the noise of  atoms
in the coherent spin state in which the atomic samples are initialized
in this paper. This does not, however, hold for higher spin.

Taking into account that the covariance matrix deals with the
transverse spin components scaled by the macroscopic longitudinal mean spin, we obtain for the corresponding
diagonal covariance matrix element:
\begin{equation}
\label{eq:gamiviaszi}
  \gamma^n_{a,ii}=(1-\eta_{\tau})^{n}\gamma^0_{a,ii}+
 \frac{\eta_{\tau}}{2}\left[\frac{1}{1-\eta_\tau}+\sum_{j=-n}^{n-2}(1-\eta_{\tau})^j\right].
\end{equation}

This analysis treats the atoms that have decayed and the ones that
have not decayed on unequal footing, and it hence breaks with the
Gaussian state Ansatz, which assumes that all information is in the
collective variance and mean values for the entire atomic ensemble.
The analysis does not make it easy to treat the coherent part of the
interactions and the measurement back action, that we are interested
in, and we hence wish to investigate, whether restoration to the
Gaussian state Ansatz will yield a large discrepancy with the exact
results. To this end, we go back to the update formula, Eq.
(\ref{eq:szdecay}), and insist that the variance obtained here should be
treated as the variance describing a Gaussian state ensemble, i.e., we
do not discriminate between the two kinds of atoms.  In subsequent
time steps, we thus simply iterate the same expression, as if all
atoms contribute evenly to the joint variance. The result of this
iteration is readily obtained:
\begin{equation}
  \label{eq:gamupdate}
  \gamma_a(t+\tau)=(1-\eta_\tau)\gamma_a(t)+
\frac{N_a}{4 J_x(t+\tau)}\eta_\tau \mathbf{I_4}.
\end{equation}
Note that since $J_x(t)=(N_a/2)e^{-\eta t}$ the last term will diverge
exponentially in time. In order to compare with Eq. (\ref{eq:gamiviaszi}) we
iterate the diagonal elements of Eq. (\ref{eq:gamupdate}) n times:
\begin{equation}
\gamma^{n}_{a,ii}=(1-\eta_{\tau})^n\gamma^{0}_{a,ii}+\eta_{\tau}\sum_{j=1}^n(1-\eta_{\tau})^{n-2j}.
\label{eq:gamiviagami}
\end{equation}

In the continuous limit $t=n\tau$ with $\eta\tau\to0$ and $n\to\infty$
both Eq. (\ref{eq:gamiviaszi}) and Eq. (\ref{eq:gamiviagami}) yield the same
behavior:
\begin{equation}
  \label{eq:gamtdecay}
  \gamma_{a,ii}(t)=e^{-\eta t}\gamma_{a,ii}(t=0)+\sinh(\eta t)
\end{equation}
This supports the use of the update formula, Eq. (\ref{eq:gamupdate}),
with the underlying assumption of a Gaussian state, together with the
evolution of $\gamma$ due to interaction and measurements. This approach
was also utilized in \cite{Klemens,magnetolbm} albeit not with the
careful justification presented above.

When light absorption is included the interaction of a light
segment with one sample $i$ is described by \cite{Klemens}:
\begin{eqnarray}
  \label{eq:gamnewloss}
\gamma(t+\tau)&=&
\bar{D}(\eta_\tau,\epsilon)S_i(\kappa_\tau)\gamma(t)
S_i(\kappa_\tau)^T \bar{D}(\eta_\tau,\epsilon)\nonumber\\&&+
D_i(\eta_\tau,\epsilon)\gamma_{\mathrm{noise},i}
\end{eqnarray}
where, following the above argument, for the first sample we have,
$D_1(\eta_1,\epsilon_1)=\mathrm{diag}(\eta_1,\eta_1,0,0,\epsilon_1,\epsilon_1)$,
$\bar{D}_1(\eta_1,\epsilon_1)=\sqrt{1-D_1(\eta_1,\epsilon_1)}$ and
$\gamma_{\mathrm{noise},1}=\mathrm{diag}(\xi,\xi,0,0,1,1)$ (and
similarly $D_2(\eta_2,\epsilon_2)=\mathrm{diag}(0,0,\eta_2,\eta_2,
\epsilon_2,\epsilon_2)$,
$\gamma_{\mathrm{noise},2}=\mathrm{diag}(0,0,\xi,\xi,1,1)$).  The
factor $\xi\equiv N_{at}/\left<J_x(t)\right>$ starts out as 2 and
increases exponentially because of the decay of the mean spin due to
excitation and subsequent decay of atoms.

As in section \ref{sec:nonoise} the differential equation for the
correlation matrix can be found:
\begin{equation}
  \label{eq:ricnoise}
   \frac{d\gamma_a}{dt} = \mathbf{\tilde{r}} \gamma + \gamma\mathbf{\tilde{r}}^T
+\tilde{A}-(1-\epsilon)\kst\gamma \tilde{B} \gamma^T
\end{equation}
where
\begin{equation}
  \label{eq:A}
  \tilde{A}=\left( \begin{array}{cccc}
\kst +\xi(t)\eta& 0 & \kst\sqrt{1-\epsilon} & 0  \\
 0 & \xi(t)\eta & 0 & 0  \\
 \kst\sqrt{1-\epsilon} & 0 & \kst+\xi(t)\eta & 0  \\
 0 & 0 & 0 & \xi(t)\eta
\end{array} \right)
\end{equation}
\begin{equation}
  \label{eq:B}
  \tilde{B}=\left( \begin{array}{cccc}
0 & 0 & 0 & 0  \\
 0 & 1-\epsilon & 0 &  \sqrt{1-\epsilon} \\
 0 & 0 & 0 & 0  \\
 0 & \sqrt{1-\epsilon} & 0 & 1
\end{array} \right)
\end{equation}
and $\mathbf{\tilde{r}}= \mathbf{r}-(\eta/2) \mathbf{I_4}$, where $
\mathbf{r}$ is defined in Eq. (\ref{eq:r}). Note that the
coefficients are now time dependent which complicates matters
slightly.

This Ricatti equation can of course easily be solved numerically but
is in its most general form too complicated to admit analytical solution.
If noise terms arising from the absorption of light, i.e. all terms
involving $\epsilon$ are neglected, the Ricatti equation can be solved
without rotations and in the strongly rotated regime. These solutions will be
derived first and then the general results will be discussed.

\subsection{Analytical results}
\subsubsection{Without rotations}
\label{sec:analnorot}

Without rotations $\gamma_a$ becomes diagonal in the sum/difference
basis if we neglect the photon absorption ($\epsilon=0$). The two
components involved in the measurement give: {\setlength\arraycolsep{1
    pt}
\begin{eqnarray}
  \label{eq:gammanoise}
  \gamma_{11}^{sd}&=&\textrm{Var}(x_{A1}+x_{A2})=e^{-\eta t}[e^{2 \eta t}+2\alpha_0\eta t]\\
\gamma_{33}^{sd}&=&\textrm{Var}(p_{A1}+p_{A2})\nonumber\\
&=&\frac{e^{\eta t }\,\left( \delta
       \cosh (\delta\eta t)
 + \sinh (\delta\eta t )
      \right) }{{\delta}
\cosh (\delta\eta t) +
    \left( 1 + 2\,\alpha_0  \right) \,\sinh (\delta\eta t )}
\label{eq:gam33}
\end{eqnarray}}
where $\delta=\sqrt{1+4\alpha_0}$. The two remaining components
increase exponentially:
\begin{equation}
  \label{eq:gam22}
\gamma_{22}^{sd}=\textrm{Var}(x_{A1}-x_{A2})=\gamma_{44}^{sd}=\textrm{Var}(p_{A1}-p_{A2})=e^{\eta
t}
\end{equation}

\begin{figure}[ht]\centering
   \includegraphics[angle=0,width=0.4\textwidth]{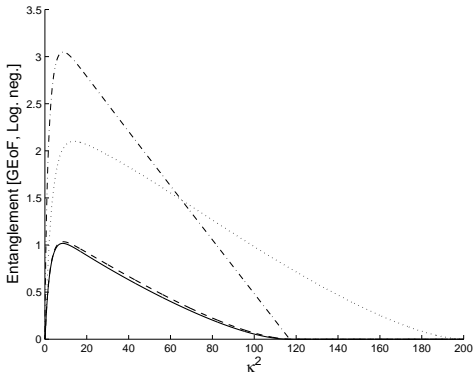}
  \caption{\small
    The generated entanglement vs. time of numerical simulation (full)
    and analytical solution(dashed). The optical density is 100 giving
    rise to only a slight difference between the two curves. The
    numerical solution is also represented in terms of the logarithmic
    negativity (dash-dot) which will be discussed in a later section.
    For further comparison a numerical simulation of GEoF with
    rotations ($\theta_{\mathrm{total}}=100\cdot2\pi$) is included (dotted).}
\label{fig:entvstime}
\end{figure}

The validity of this solution can be tested by comparing with full
numerical solutions. In Fig. \ref{fig:entvstime} we plot the GEoF as
a function of $\kappa_t^2$ (proportional with the total number of
transmitted photons) and we see that in the case of dissipation, the
entanglement reaches a maximum, and hereafter it decays and vanishes
at a point when -  because of the decay of the macroscopic spin - the
atomic samples are in a mixed state with too few correlations to
display actual entanglement.

As can be seen in Fig. \ref{fig:entvstime} the analytic
expression fits very well with the nontrivial result of the
numerical calculation. The comparison is made for an optical
depth, $\alpha_0=100$, corresponding to
$\epsilon=\frac{\Gamma^2}{\Delta^2}\cdot \alpha_0=0.0025 \ll1$. For
larger optical depths, the quality of the analytical expression
deteriorates whereas for lower optical densities the error of the
analytical expression is negligible. This thus establishes the
regime of validity for the analytic expression. 

It is interesting to investigate when the maximum entanglement is
reached and at what level as a function of $\alpha_0$ and $\eta$. By
differentiating $\Delta=\sqrt{\gamma_{22}\gamma_{33}}$ with respect to
time we get:
\begin{equation}
  \label{eq:tcrit}
  t_{\mathrm{crit}}=\frac{\mathrm{arccosh}(\frac{1}{2}{\sqrt{\frac{-2\,{\alpha }^2 - 4\,{\alpha }^3
+{\left( 1 + 5\,\alpha  + 4\,{\alpha }^2 \right) }{\sqrt{{\alpha }^3\,}}}{
              {\alpha }^3}}})}{{\sqrt{1 + 4\,\alpha }}\,\eta }
\end{equation}
Note the simple inverse scaling with $\eta$. Since $\eta$ and t
only appear as a product in Eqs. (\ref{eq:gammanoise}) and
(\ref{eq:gam22}) the inverse scaling of $t_{\mathrm{crit}}$ with $\eta$ means
that the maximum achievable entanglement will be independent of
$\eta$. This behavior could also be predicted from the fact that
as seen in Eqs. (\ref{eq:gam33}) and (\ref{eq:gam22}) $\Delta$ is a
function of the product $\eta t$.

\begin{comment}

The analytic expression for the maximum achievable entanglement is complicated
but computable.
\begin{figure}[ht]\centering
  \includegraphics{maxent_anal_vs_numeric2.eps}
  \caption{\small
Maximum achievable entanglement vs. optical density
}
\label{fig:maxent}
\end{figure}
In Fig. \ref{fig:maxent} this is plotted together with the numerical
results.  The agreement is very nice and again confirms the validity
of our analytic expression within this regime.

\end{comment}

\subsubsection{Many rotations}

To solve Eq. (\ref{eq:ricnoise}) in the case of many rotations
we define a new set of canonical operators \cite{distent}:
\begin{eqnarray}
\begin{array}{cc}
\hat{x}_{A}=\frac{\jh'_{y1}-\jh'_{y2}}{\sqrt{2J_x}},
~~~&\hat{x}_{L}=\sqrt{\frac{2}{S_x
T}}\int_0^T\sh_y(t)\cos(\omega_L t )dt\\
\hat{p}_{A}=\frac{\jh'_{z1}+\jh'_{z2}}{\sqrt{2J_x}},
~~~&\hat{p}_{L}=\sqrt{\frac{2}{S_x
T}}\int_0^T\sh_z(t)\cos(\omega_L t )dt
\end{array}
\label{eq:xplab}
\end{eqnarray}
where $\jh'_k$ refers to rotating frame coordinates, i.e.
coordinates rotated an angle $\omega_L t$ compared to the usual
lab frame coordinates.  With these collective operators we regain the same
formal description of the interaction as in the non-rotating case:
\begin{eqnarray}
\label{eq:int}
\begin{array}{cc}
\hat{x}_{A}^{\textrm{out}}=\hat{x}_{A}^{\textrm{in}}+\kappa\hat{p}_{L}^{\textrm{in}}
,\qquad&\hat{p}_{A}^{\textrm{out}}=\hat{p}_{A}^{\textrm{in}}\\
\hat{x}_{L}^{\textrm{out}}=\hat{x}_{L}^{\textrm{in}}+\kappa\hat{p}_{A}^{\textrm{in}},
\qquad&\hat{p}_{L}^{\textrm{out}}=\hat{p}_{L}^{\textrm{in}}
\end{array}
\end{eqnarray}
Note that in this formalism the atomic correlation matrix is reduced
to a 2x2 matrix whereas the correlation matrix of light remains a 2x2
matrix.

To illustrate the use of this new description first neglect all decoherence.
With the usual update equation for the system we can form the
differential equation for the atomic correlation matrix:
\begin{equation}
\frac{\partial \gamma_a}{\partial t} = \ks \left(%
\begin{array}{cc}
  1 & 0 \\
  0 & 0 \\
\end{array}%
\right) -\ks\gamma_a\left(%
\begin{array}{cc}
  0 & 0 \\
  0 & 1 \\
\end{array}%
\right) \gamma_a
\end{equation}
When this is inserted into the Ricatti equation we easily derive
the solution:
\begin{equation}
\gamma_a(t)=\left(%
\begin{array}{cc}
  1+\kappa_t^2 & 0 \\
  0 & \frac{1}{1+\kappa_t^2} \\
\end{array}%
\right)
\end{equation}
which coincides with the rotated solution found in (13).

In order to treat decoherence analytically an approximation similar to
that of section \ref{sec:analnorot}, i.e. ignoring all $\epsilon$
dependent noise contributions, has to be made. The loss of light in
the first sample, represented by $\epsilon$ will create asymmetries
between the two samples which would cause the $\hat{x}_A$, $\hat{p}_A$
formalism of Eqs.  (\ref{eq:xplab},\ref{eq:int}) to break down. The
regime of validity is thus expected to be similar to that of section
\ref{sec:analnorot}, i.e.  $\alpha_0\lesssim100$.

We will consider the case of small values of $\eta t$, i.e. the atomic
samples retain a constant polarization along the x-axis and the
differential equation has constant coefficients.

\begin{comment}
The differential equation now turns into:
\begin{equation}
\frac{\partial \gamma_a}{\partial t} = - \eta \gamma_a +\ks \left(%
\begin{array}{cc}
  1 & 0 \\
  0 & 0 \\
\end{array}%
\right) -\ks\gamma_a\left(%
\begin{array}{cc}
  0 & 0 \\
  0 & 1 \\
\end{array}%
\right) \gamma_a+\eta 2 \xi \left(%
\begin{array}{cc}
  1 & 0 \\
  0 & 1 \\
\end{array}%
\right)
\end{equation}

This gives:

\begin{equation}
\gamma_a=\left(%
\begin{array}{cc}
  e^{-\eta t}\left[(\alpha_0+2\xi)(e^{\eta t}-1)+1\right]  & 0 \\
  0 & \frac{(e^{\eta t\beta}-1)(4\xi-1)+(e^{\eta t \beta}+1)\beta}{(e^{\eta t\beta}-1)(1+2\alpha_0)+(e^{\eta t \beta}+1)\beta} \\
\end{array}%
\right)
\end{equation}
where $\beta=\sqrt{1+8\alpha_0\xi}$.

Assuming $\eta t$ to be small we get:
\begin{eqnarray}
\gamma_{11}&\approx& (1-\eta t)(1+(\alpha_0+2\xi)\eta t)\to 1+\ks t\\
\gamma_{22}&\approx& \frac{2+\eta
t(4\xi-1+\sqrt{1+8\alpha_0\xi})}{2+\eta
t(1+2\alpha_0+\sqrt{1+8\alpha_0\xi})}\to\frac{1}{1+\ks t}\label{eq:gam22rot}
\end{eqnarray}
\end{comment}

\begin{equation}
\gamma_a=\left(%
\begin{array}{cc}
  e^{-\eta t}\left[(\alpha_0+4)(e^{\eta t}-1)+1\right]  & 0 \\
  0 & \frac{7(e^{\eta t\beta}-1)+(e^{\eta t \beta}+1)\beta}{(e^{\eta t\beta}-1)(1+2\alpha_0)+(e^{\eta t \beta}+1)\beta} \\
\end{array}%
\right)
\end{equation}
where $\beta=\sqrt{1+16\alpha_0}$.

An expansion in small $\eta t$ yields:
\begin{eqnarray}
\gamma_{11}&\approx& (1-\eta t)(1+(\alpha_0+4)\eta t)\\
\gamma_{22}&\approx& \frac{2+\eta
t(7+\sqrt{1+16\alpha_0})}{2+\eta
t(1+2\alpha_0+\sqrt{1+16\alpha_0})}\label{eq:gam22rot}
\end{eqnarray}
Eq. (\ref{eq:gam22rot}) shows that a value of $1+2\alpha_0>7$, is required
in order to get squeezing in $p_A$. Again we see that the
variances are only functions of the two variables $\alpha_0$ and
$\eta t$, and in the limit of vanishing noise we obtain
$\gamma_{11} \rightarrow 1+\kappa_t^2$ and $\gamma_{22}
\rightarrow 1/(1+\kappa_t^2)$, in accordance with our earlier
results.

\subsection{General results}

We now turn to general numerical results beyond the regime of
validity of the analytical solutions. First however we need to
note that the GEoF is not applicable in the presence of the large
asymmetries between the two cells that occurs at high $\alpha_0$. We
therefore introduce the logarithmic negativity of
\cite{Audenaert}.

\subsubsection{Logarithmic negativity}
\begin{figure}[t]\centering
  \includegraphics[width=0.35\textwidth]{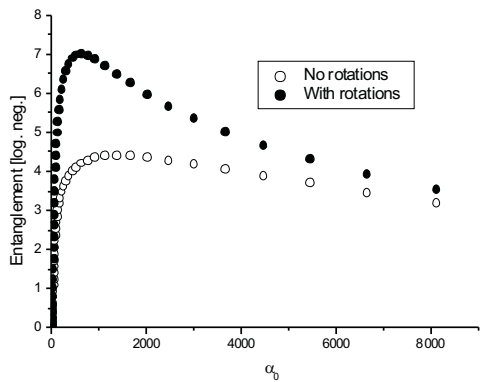}
  \caption{\small
    The maximum generated degree of entanglement as a function of
    $\alpha_0$ with and without rotations.  }
\label{fig:maxentwwoutrot}
\end{figure}

In this measure the symplectic spectrum of the partial transpose of
the covariance matrix $\gamma^{T_A}$ is used to calculate the degree
of entanglement. In the partial transposition all covariance matrix
elements involving $p_A$ and other observables are multiplied by (-1).
The symplectic eigenvalues are most conveniently calculated by
computing the eigenvalues of the matrix $\sigma^{-1}\gamma$, where
$\sigma$ is the matrix specifying the commutators:
$\sigma_{\alpha\beta}=-i[y_\alpha,y_\beta]$. For m systems this gives
2m complex eigenvalues $\lambda_k$. The logarithmic negativity
entanglement measure is then given by:
\begin{equation}
  \label{eq:logneg}
  \mathrm{log.neg.}=-\sum_{k=1}^{2m}\log_2[\mathrm{min}(1,2|\lambda_k|)]
\end{equation}
This is a valid entanglement monotone even for asymmetric samples
but it does not coincide with the Von Neuman entropy for pure
states as the GEoF does. If $\gamma$ is diagonal in the
sum/difference basis the eigenvalues can be expressed in terms of
the diagonal entries:
\begin{equation}
\lambda=\pm i \sqrt{\frac{\gsd_{11}\gsd_{44}+\gsd_{22}\gsd_{33}
\pm(\gsd_{11}\gsd_{44}-\gsd_{22}\gsd_{33})}{8}}
\end{equation}
Without losses the covariances are specified by Eq.
(\ref{eq:a+-}) giving a logarithmic negativity of
$\log_2(\kappa_t^2)+1$ and $2\log_2(\kappa_t^2)$ in the
non-rotated and the rotated regimes respectively at
$\kappa_t^2\gg1$. This is about a factor of two larger than the
GEoF as specified in Eq. (\ref{eq:geoflimit}). The difference
between GEoF and the logarithmic negativity in the presence of
decoherence is illustrated in Fig. \ref{fig:entvstime} where
log. neg.(dash-dotted) is seen to be approximately a factor of
three larger than the GEoF (full drawn).

\subsubsection{Numerical results}
In Fig. \ref{fig:maxentwwoutrot} we see the degree of entanglement
generated with and without rotations. For low optical densities the
improvement in the degree of entanglement due to rotations of the
samples is close to the factor of two, which we got in the absence of
decoherence. This is also illustrated in Fig. \ref{fig:entvstime} for
the case of $\alpha_0=100$. For higher optical densities the benefits
of rotations seem to decrease. This is probably the result of the
complicated interplay between the beneficial rotations and the
increasing asymmetry between the two samples at high optical
densities. It would certainly be a of great interest to study this
further and to consider protocols where the samples are illuminated
with lasers from both sides to restore their symmetry.

\begin{figure}[t]\centering
  \includegraphics[width=0.33\textwidth]{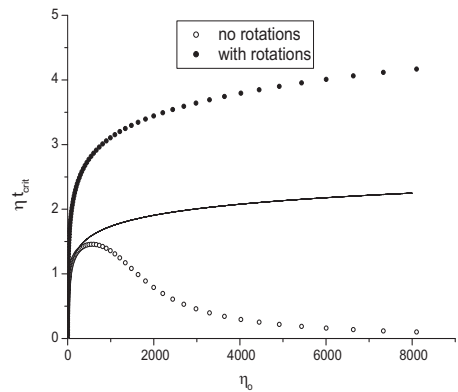}
  \caption{\small
    The $\eta t$ at which the entanglement has decayed to zero with
    and without rotations. The solid curve is calculated from Eqs.
    (\ref{eq:gammanoise}-\ref{eq:gam22}) assuming no rotations and no
    asymmetry between the two samples. }
\label{fig:noent}
\end{figure}
In Fig. \ref{fig:noent} we show at which accumulated $\eta t$ the
entanglement has decreased to zero. The theoretical curve is
calculated from Eqs. (\ref{eq:gammanoise}) to (\ref{eq:gam22}). This
solution fits the numerical results perfectly within the regime of
validity, that is up to $\alpha=200$. After this point entanglement
decays rapidly, indicating that light losses which
introduce asymmetry between the two samples are detrimental to the
entanglement.  As can be seen, the entanglement is not only increased
by rotations. The rotated samples also seem to be a lot less
vulnerable to light losses. Note that the decay in all cases happens
in the vicinity of $\eta t=1$ which could be expected from the
exponential character of the decay of the mean spin and the resulting
exponential growth of $\gamma_{\mathrm{noise}}$.

\section{Mean values}
\label{sec:meanvalues}
The generated entanglement is of no practical use unless the mean
values of the atomic parameters are known.  These will be affected
by the interaction according to:
\begin{equation}
  \label{eq:meanint}
  \left<\mathbf{y}(t+\tau)\right>=\bar{D_2}S_2\bar{D_1}S_1
\left<\mathbf{y}(t)\right>
\end{equation}

The mean values of the atomic variables after the interaction
$\left<\mathbf{y_a(t+\tau)}\right>$ will subsequently transform
according to the result of the homodyne measurement as:
\begin{equation}
  \label{eq:meanhomo}
  \left<\mathbf{y}_a(t+\tau)\right>\to \left<\mathbf{y}_a(t)\right> +
\gamma_c (\pi\gamma_b\pi)^{-1}(\chi,0)^T
\end{equation}
where $\chi$ is the difference between the measurement result and the
mean value of the detected field component. That is, the evolution of
the mean values is determined by the value of the correlation matrix
at a given time.  Note also that it is the deviation from the mean
value of the light that determines the evolution of the atomic mean
values and not the measurement result itself. This will be discussed
further below.  The correlation matrix can be made diagonal in the
sum/difference basis.  We can therefore without loss of generality
assume $\gamma_a(t)$ to be diagonal.

Having established the relatively large domain of validity of the
analytical solution of the lossy system without rotations and
negligible absorption losses in section \ref{sec:analnorot} we
now apply the results of that section to Eq.
(\ref{eq:meanhomo}). Given the fact that all mean values start out
as zero we obtain the transformation of the atomic variables:
\begin{equation}
  \label{eq:meaninthomo}
  \left<\mathbf{y}_a(t+\tau)\right>=(1-\eta_\tau /2)
\left<\mathbf{y}_a(t)\right> + \kappa_\tau
  \gamma_{33}^{sd}\left[
\begin{array}{c}
0\\
1\\
0\\
1
\end{array}
\right] \chi
\end{equation}
where $\gamma_{33}^{sd}$ has the value specified in Eq.
(\ref{eq:gam33}). We see that the two x-variables will have zero
mean at all times and the two p-variables will experience the same
evolution.

After the interaction we get in the limit of vanishing $\epsilon$:
\begin{equation}
  \label{eq:gam55}
 \gamma_{55}=2\textrm{Var}(x_L)=1+2 \gamma_{33}^{sd}\kappa^2_\tau
\end{equation}

Since we always work in the regime $\kappa^2_\tau\ll 1$ and
$\gamma_{33}^{sd}\approx1$ in the relevant regime we can safely assume
$\gamma_{55}=1$. This means that the difference between the result of
the homodyne measurement of $x_L$ and its expectation value, will be a
Gaussian random variable, $\chi$ with zero mean and variance 1/2.

\begin{comment}

One can now either try to solve the recursion Eq.
(\ref{eq:meaninthomo}) or transform it into a differential equation
and then try to solve this. We start with the first option. Naming
the non-stochastic part of the last term in Eq.
(\ref{eq:meaninthomo}) of the i'th iteration,
$z_i=\tilde{\kappa}\sqrt{\tau}\gamma_{33}^{sd}$, and $1-\eta\tau
/2=\delta$ we can solve the recursion relation to obtain:
\begin{equation}
  \label{eq:recur}
  \left<p(n\cdot\tau)\right>=\delta^n
\left<p(0)\right> +z_n \chi_ n+\delta z_{n-1}\chi_ {
n-1}+...+\delta^n z_0\chi_ { 0}
\end{equation}
That is, $\delta^{n-i}\cdot z_i$ is the weight given to the i'th
stochastic result. Since we assume the atomic variables to have
zero mean initially the first term vanishes. Note also that
$\delta^n\approx e^{- \eta n \tau/2}$ implies that the mean value
has a ''memory'' stretching approximately
$t_{\mathrm{mem}}=2/\eta$ back in time. We now make the assumption
that the acquired mean value of the light during the interaction
time, $\tau$, is very small compared to the standard deviation of
the light. This will be shown to be valid below. $\chi$ will now
essentially be the measurement record and Eq. (\ref{eq:recur}) tells
us how to calculate the mean values given the measurement record.
Alternatively
\end{comment}

Eq. (\ref{eq:meaninthomo}) can be transformed into a stochastic
differential equation with the solution:
\begin{equation}
  \label{eq:meananal}
  \left<p(T)\right>=
\int_{0}^T e^{-\eta (T-t) /2} \frac{\tilde{\kappa}}{\sqrt{2}}
\gamma_{33}^{sd}dW
\end{equation}
where $dW= \sqrt{2 dt}\chi$ is a stochastic Wiener increment with zero
mean and variance dt. We recognize an exponential memory decay for
early detection events, with $t_{mem}\approx 2/\eta$, and a covariance
matrix weight factor, $\gamma^{sd}_{33}(t)$, based on the state of the
atoms at the particular time. From Eq (\ref{eq:meananal}) it follows
that the conditional mean value of $p$ will be a stochastic
variable with zero mean and variance:
\begin{equation}
  \label{eq:varmean}
  \textrm{Var}(\left<p(T)\right>)=
\frac{\tilde{\kappa}^2}{2}\int_{0}^T e^{-\eta (T-t)
}(\gamma_{33}^{sd})^2 dt
\end{equation}
%\begin{figure}[t]
%\includegraphics[width=0.5\textwidth]{varmeanp_anal_divideby2.eps}
%\label{fig:varmean} \caption{\footnotesize The variance of the mean.}
%\end{figure}

%A numerical integration of this equation shows the variance to be
%in the vicinity of unity  for all relevant values of $\eta t$ and
%since Eq. (\ref{eq:meanlight}) shows us that the mean of p must be
%multiplied with the factor $\kappa_\tau\ll1$ the
%noise in the light will dominate the acquired mean completely,
%thus validating the approximation of using the measured field
%quadrature in place of $\chi$ in Eq. (\ref{eq:meanhomo}).

%To relate these results to those of \cite{phd},
% neglect decoherence and rotations giving
% $\gamma_{33}^{sd}=1/(1+2\ks_t)$. The mean value after one iteration is
% given by:
%\begin{equation}
%  \label{eq:meansimple}
%  \left<p_1\right>=\frac{\kappa_\tau}{1+2\ks_\tau}\chi
%\end{equation}
%which is in exact accordance with the results of the simple pure
%state calculations of \cite{phd}. 

To illustrate the physical interpretation of Eq. (\ref{eq:varmean})
we neglect decoherence and rotations.  In this simple case we can
perform the integration in Eq. (\ref{eq:varmean}) analytically:
\begin{equation}
  \label{eq:varmeansimple}
  \textrm{Var}(\left<p(T)\right>)=
\frac{\tilde{\kappa}^2}{2}\int_{0}^T
\frac{1}{(1+2\tilde{\kappa}^2t)^2}dt=\frac{1}{2}\frac{\tilde{\kappa}^2T}{1+2\tilde{\kappa}^2T}
\end{equation}
Since $\left<p_1(T)\right>=\left<p_2(T)\right>$,
$\textrm{Var}(\left<p_1(T)\right>+\left<p_2(T)\right>)$ will be four
times the result of Eq. (\ref{eq:varmeansimple}). It follows that at
any time
\begin{equation}
 \gamma_{33}^{sd}+ 4
 \textrm{Var}(\left<p\right>)=\frac{1}{1+2\ks_t}+\frac{2\ks_t}{1+2\ks_t}=1
\end{equation}
Initially, the expectation value is well determined (0) whereas the
quantum deviation from this value is given by the initial Gaussian
distribution. After a significant interaction time the quantum
mechanical uncertainty will be reduced but the value within the
initial distribution at which the expectation value settles is
uncertain.

\begin{comment}
The form of Eq. (\ref{eq:meananal}) might seem to suggest that we
should apply different weights to measurements at different times
according to the value of $\gamma_{33}^{sd}$. This would lead to the
fairly surprising conclusion that the optimum feedback strategy would
involve a time dependent feedback strength. First however, we need to
note that $\chi$ is the measured deviation from the expected mean
value and not the measurement result itself. This leads to a subtle
but very important difference. To see this consider the case of no
decoherence and no rotations and imagine that we have a measurement
record discretized over $n$ small time bins of duration $\tau$. The
mean value of the $p$ variables after time $n\tau$ is given by:
\begin{eqnarray}
  \label{eq:recursivemean}
  \left<p(n\tau)\right>&=& \left<p(0)\right>+
\sum_{i=0}^n\kappa_\tau\gamma_{33,i}^{sd}\chi_i\nonumber\\
&=& \left<p(0)\right>+\sum_{i=0}^n\frac{\kappa_\tau}{1+i\kappa_\tau^2}
(\tilde{\chi_i}-\left<p((i-1)\tau)\right>)\nonumber\\
&=& \left<p(0)\right>+\frac{n\kappa_\tau}{1+n\kappa_\tau^2}
\sum_{i=0}^n\frac{\tilde{\chi}}{n},
\end{eqnarray}
where $\tilde{\chi}_i$ denotes the measurement result in time bin $i$.
The last equality can be shown inductively. Equation
(\ref{eq:recursivemean}) clearly shows
\end{comment}

Eqs. (\ref{eq:meaninthomo},\ref{eq:meananal}) express the conditional
mean value of the atomic variable in terms of the difference between
the optical read-out and its expectation value. It is interesting to
obtain similar expressions in terms of the actual read-out. Since the
coherent light initially has zero mean Eq.  (\ref{eq:meanint}) shows
us that the measured light component will have the mean value:
\begin{equation}
\label{eq:meanlight}
\left<x_L(t)\right>=\kappa_\tau(\left<p_1(t)\right>+\left<p_2(t)\right>)
\end{equation}
in the absence of $\epsilon$ decay.  We thus start by writing $\chi$
in Eq. (\ref{eq:meaninthomo}) as $\tilde{\chi}-2
\kappa_\tau\left<p(t)\right>$. The atomic $p$ variable thus changes as
\begin{equation}
  \label{eq:pmeanrecur}
  \left<p(t+\tau)\right>=(1-\eta_\tau/2)\left<p(t)\right>+
\kappa_\tau\gamma_{33}^{sd}(t)(\tilde{\chi}-2 \kappa_\tau\left<p(t)\right>)
\end{equation}
where $\tilde{\chi}$ is the random detector output. Taking the limit
of inifitesimal $\tau=dt$ and defining the measured Wiener increment
$d\tilde{W}=\sqrt{2dt}\tilde{\chi}$, we can integrate Eq.
(\ref{eq:pmeanrecur}):
\begin{equation}
  \label{eq:pmean}
  \left<p(T)\right>=\int_{0}^T e^{-\eta (T-t) /2-\int_t^T2\tilde{\kappa}^2\gamma_{33}^{sd}(t')dt'} \frac{\tilde{\kappa}}{\sqrt{2}}
\gamma_{33}^{sd}(t)d\tilde{W}
\end{equation}
In the limit of $\eta=0$, $\gamma_{33}^{sd}(t)=1/(1+2\kst t)$, and we
can explicitly integrate the argument of the exponential function in
Eq. (\ref{eq:pmean}), and we obtain
\begin{eqnarray}
  \label{eq:pmeansimple}
  \left<p(T)\right>&=&\int_{0}^T \frac{1+2\kst t}{1+2 \kst T}
 \frac{\tilde{\kappa}}{\sqrt{2}}\gamma_{33}^{sd}d\tilde{W}\nonumber\\
&=&\frac{\tilde{\kappa}}{\sqrt{2}(1+2\kst T)}\int_{0}^T 
 d\tilde{W}
\end{eqnarray}
This remarkable result shows  that all measurements should be
weighted equally in the absence of decoherence and rotations. When
transformed back into regular angular momenta the common weight factor
has a clear interpretation as the ratio between the shot to shot
atomic noise contribution, the so-called projection noise, and the
total noise in complete accordance with the result obtained in
\cite{distent}. 

We stress that the result (\ref{eq:pmeansimple}) was obtained for the
non-rotated and  non-decaying atomic systems. Decay can be included
easily according to Eq. (\ref{eq:pmean}). In addition to the
exponential damping term, this will involve a more complicated
expression for $\gamma_{33}(t)$, and the conditioned mean value of
p(T) will no longer be given by the integrated measurement outcome. We
can understand that atomic mean values attained early during the
measurements decay and hence they contribute less to the final
$\langle p(T)\rangle$ than the most recent contributions to the
optical detection. Since decay is inevitable, our analysis suggests
that experiments must be carried out so that the optical signal is
recorded in time bins which are much shorter than the atomic
decoherence time $\eta^{-1}$. The results will then be in good
agreement with our continuous update theory. 

 
\section{Conclusion}
\label{sec:conclusion}

In conclusion, we have presented a theory for the preparation of
entangled atomic ensembles by detection of the Faraday polarization
rotation of a continuous optical field passing though both ensembles.
Our Gaussian Ansatz is very well justified for large atomic ensembles
and for free space atom-light interaction, where only the interaction
with many photons appreciably modifies the atomic state. The theory
incorporates the interaction between the atoms and the optical field,
atomic decay, and the measurement induced transformation of the atomic
state. The reduction of the full quantum state description to a simple
Gaussian state fully represented by a set of mean values and a
covariance matrix makes the system straight forward to deal with numerically, 
and analytical results can be obtained in several  important cases.

The entanglement between the atomic ensembles is quantified by the
Gaussian Entanglement of Formation and the Logarithmic Negativity, and
we identify the optimal performance of the entanglement scheme in the
presence of atomic decay.  Our analysis confirms a number of results,
derived in less complete or purely numerical studies, and it presents
an intuitive physical picture of the continuous transformation of the
atomic quantum state from an initial state with no atomic correlations
into a state with stronger correlations of the quantum observables,
around mean values with a broader random distribution - but known by
the experimentalist in every implementation of the experiment.

The results of our analysis are relevant for current experimental
efforts to exploit entangled atomic ensembles for quantum  purposes,
but we also wish to emphasize the strengths of the theoretical method,
which make it readily adapt to a wide variety of experiments.

\acknowledgments{Discussions with Lars
  Bojer Madsen, Eugene Polzik, Brian Julsgaard, Jörg Helge Müller, and
  Anders Sørensen are gratefully acknowledged}

\bibliographystyle{apsrev}
\bibliography{paper}
\end{document}